\documentclass[preprint]{aastex}
\usepackage[utf8]{inputenc}
\usepackage{amsmath}
\usepackage{amsfonts}
\usepackage{amssymb}
\usepackage{graphicx}
\usepackage{natbib}

\begin{document}

\title{Convergence and shear statistics in galaxy clusters as a result of Monte Carlo simulations} 
\author{Alexander Poplavsky} 
\affil{Department of Theoretical Physics and Astrophysics, \\ Belarusian State University} 
\affil{4 Nezavisimosti Avenue, Minsk 220030, Belarus} 
\email{a.poplavsky@mail.by}

\begin{abstract}
In this paper the influence of galaxy cluster halo environment on the deflection properties of its galaxies is investigated. For this purpose circular and elliptical projected cluster haloes obeying Einasto density profiles are modelled in the $\Lambda$CDM cosmological model. By Monte-Carlo simulations external shear and convergence are calculated for random positions of a test galaxy within its cluster. Throughout the simulations the total virial mass, profile concentration and slope parameters are varied both for cluster and its  galaxies. The cluster is composed  of smooth matter distribution (intergalactic gas and dark matter) and randomly placed galaxies. As a result of multiple simulation runs robust statistical estimations of external shear and convergence are derived for variable cluster characteristics and and its redshift. In addition, the models for external shear and convergence are applied for the galaxy lens seen through the cluster IRC-0218.  
\end{abstract}


\section{Introduction}

Gravitational lensing is one of a few available tool for probing Cold Dark Matter (CDM) in haloes of galaxies and clusters. Throughout the latest two decades a significant progress in this field has been made both in observational and theoretical astronomy \citep[see review]{bartelmann2010}. Observers have found the evidence for gravitational imprints of CDM in clusters and individual galaxies, usually investigated by means of multiple lensed images. Combining these data with theoretical and numerical models of CDM distribution, for some clusters and galaxies CDM distribution constraints have been obtained \citep{hoekstra2013}. 

In the previous decade much effort was put into finding direct or indirect evidence for CDM hierarchy scenario revealed in large-scale N-body simulations \citep{madau2008}. According to this paradigm CDM has to demonstrate strong hierarchy at all scales from galaxy clusters to galaxies and even downwards. In such a scenario a halo halo should consist of numbers of subhaloes. Given that proving the existence of such substructures is one of the most challenging astrophysical task nowadays, strong gravitational lensing methods and sensitivity of the state-of-the-art instruments (VLTI, EVN, VLBA, ATNF and upcoming SKA) are in most cases sufficient for resolving tiny effects produced by dwarf-galaxy-sized dark companions of normal galaxies. In order to fit observational results properly precise models of lensing are strongly demanded. Necessary requirements for them include thorough  accounting of additional effects that could affect lensing by substructures, including external shear and convergence by cluster mass  and effects of propagation of light rays between a source and an observer. 

Lensing by substructures has been investigating by some authors. \citet{metcalf2001} and \citet{metcalf2012} have studied the effects of substructure influence on lensed image flux ratios. Recent observations have shed light on the fact that both cluster and galactic haloes are better modelled by triaxial Einasto mass distribution \citep{kneib2011} and external shear could could contribute to the lensing by galaxies.  However some authors still apply singular isothermal spherical and elliptical (SIS and SIE) CDM halo profiles and did not include shear related to external cluster mass distribution. 


Both observational \citep{kneib2011} and numerical results today tend to generalised Navarro-Frenk-White (NFW)  \citep{munoz2001,keeton2001b}  profile for dark haloes and Einasto \citep{merritt2006,dhar2010} mass distribution for either cluster halo or its  galaxies. In addition, the external deflection effects caused by galactic clusters are in most cases ignored or estimated only roughly. The purpose of this paper is to develop precise model of a galaxy cluster halo having circular and elliptical projected surface density and Einasto spatial profile and simulate external shear and convergence for a range of parameters consistent with other cosmological simulations and $\Lambda$CDM model. 

Recently a strong gravitational lens has been discovered in the galaxy cluster IRC-0218 \citep{pierre2012,wong2014}. \citet{wong2014} have estimated the contribution of the cluster in the effective deflecting mass of the lens galaxy. In this paper I apply the numeric cluster model to obtain more comprehensive results for cluster influence. 

Angular size distances are calculated based on the recent cosmological parameters derived from the Plank best fit results 2013 \citep{plank2013}: $\Omega_\Lambda=0.6825$, $\Omega_m=0.3172$, $H_0=67.11$~km/s/Mpc. However, for modelling IRC-0218 galaxy cluster in order to compare results with other authors the model was the following: $\Omega_\Lambda=0.7$, $\Omega_m=0.3$, $H_0=70$~km/s/Mpc. Except the results of redshift dependence and IRC-0218 cluster, redshifts of a source and a deflector are always $z_s=1.5$ and $z_d=0.5$.

\section{Methods}
\subsection{Lensing quantities}

Following a standard gravitational lensing formalism, the lensing equation for a deflector in an external perturbation field reads:

\begin{eqnarray} \label{lenseq}
\mathbf{y}=\boldsymbol{\alpha}_{ext}-\left( \begin{array}{cc}
1-\kappa_{ext}-\gamma_{1ext} & -\gamma_{2ext}  \\
-\gamma_{2ext} & 1-\kappa_{ext}+\gamma_{1ext} 
\end{array} \right)\mathbf{x} + \nonumber 
\mathbf{x}-\boldsymbol{\alpha}(\mathbf{x}),
\end{eqnarray}  
where $\boldsymbol{\alpha}_{ext}$ is the external deflection acting on the centre of a galaxy in a cluster, $\boldsymbol{\alpha}$~--- the intrinsic deflection by test galaxy,
$\kappa_{ext}$ is the external convergence or dimensionless surface mass density, $\gamma_{1ext}$ and $\gamma_{2ext}$ are the components of the external shear. The latter quantities could be combined as $\gamma_{ext}=\sqrt{\gamma_{1ext}^2+\gamma_{2ext}^2}$. Hereafter in the paper I will deal only with the external deflection. All variables in (\ref{lenseq}) are dimensionless and scaled by characteristic cluster radius to angular size distance to the deflector ratio $r_s/D_d$, where $r_s$ is defined as $r_{vir}/C$, where $r_{vir}$~--- virial radius, $C$~--- concentration parameter (see next subsection). 

For modelling mass distribution within galactic clusters the Einasto density profiles is used \citep{merritt2006}: 
\begin{equation}\label{einasto}
\rho=\rho_0 \exp\left[-\frac{2}{\beta}\left(\frac{r}{r_{-2}}\right)^\beta\right].
\end{equation} 
This profile has more free parameters than NFW, so it could be better adjusted to the data and has no singularity in the centre. The recent analysis of the applications of (\ref{einasto}) can be found in \citet{dhar2010}.   
 
In a spherically symmetric Einasto model the radial deflection angle is expressed as
\begin{equation}\label{alpha_circ}
\alpha (r) = \dfrac{4 G M D_d D_{ds}}{r r_s^2 c^2 D_s}\dfrac{\mathcal{\gamma} (3/\beta, 2 r^\beta/\beta)}{\mathcal{\gamma}(3/\beta,2 {r_{out}}^\beta/\beta)},
\end{equation}
where $M$ is a virial cluster mass; $D_s$ and $D_{ds}$ are angular size distances from the source to the observer and between the deflector and the source respectively; scale radius $r_s$ is assumed to be equal to $r_{-2}$ in \ref{einasto} and is dimensional; $\mathcal{\gamma}$ is the lower incomplete gamma function. The expression (\ref{alpha_circ}) will be then transformed to the deflection of elliptical projected mass using proper variable transformation.  The similar approach has been adopted in \citet{meneghetti2003} that differs from direct calculation  of elliptical mass defined by \citet{retana-montenegro2012}. 

The shear and convergence are defined in terms of second derivatives of the deflection:
\begin{eqnarray}
\kappa_{ext}=\frac{1}{2}\left(\alpha_{11}+\alpha_{22}\right), \nonumber \\
\gamma_{1ext} = \frac{1}{2}\left(\alpha_{11} - \alpha_{22}\right), \nonumber \\
\gamma_{2ext} = \alpha_{12},
\end{eqnarray}
where $\alpha_{ij}\equiv \partial \alpha_i / \partial x_j$.

Is the case of elliptical symmetry a family of ellipses in the lens plane is defined as follows:
\begin{equation}
\xi^2=x_1^2q+x_2^2/q,
\end{equation} 
where $q$ is the projected axial ratio, or ellipticity. Such a definition preserves the mass enclosed within $\xi$ when $q$ varies. In terms of this variable the derivatives $\boldsymbol{\alpha}(\xi)$ are
\begin{eqnarray}\label{derivatives}
\alpha_{11}=\frac{\partial\alpha(\xi)}{\partial \xi}\frac{x_1^2 q^2}{\xi^2}+\alpha(\xi) \left[\frac{q}{\xi}-\frac{x_1^2 q^2}{\xi^3}\right], \nonumber \\
\alpha_{22}=\frac{\partial\alpha(\xi)}{\partial \xi}\frac{x_2^2}{q^2 \xi^2}+\alpha(\xi) \left[\frac{1}{q \xi}-\frac{x_2^2}{q^2\xi^3}\right], \nonumber \\
\alpha_{12}=\alpha_{21}=\frac{\partial \alpha(\xi)}{\partial\xi}\frac{x_1 x_2}{\xi^2}-\alpha(\xi)\frac{x_1 x_2}{\xi^3}. 
\end{eqnarray}

\subsection{Cluster model}

In this subsection I describe the numerical model of the shear and convergence in a galaxy cluster. For numerical simulations the cumulative contribution of all galaxies are set as $m_{gal}=0.1M$ according to data in \cite{natarajan1998}.  The rest mass is treated as smooth component. The outer halo radius is related to the scale radius as $r_{vir}=Cr_{-2}$ and defined as a radius at which $\rho= \rho_{200}$, where $\rho_{200}=200\rho_{crit}$. This is done in order to obtain results consistent with the accepted paradigm of $\Lambda$CDM haloes. 

Cluster mass is varied obeying the mass distribution function constructed here to fit numerical results from Bolshoi simulations \citep{klypin2011}.   Data from \citet{klypin2011} are fitted with the mass function based on which the following random mass generator is constructed:
\begin{equation}\label{mass_function}
M_{12} = M_{min 12} + \left[\frac{\log(1.0+u)}{a}-\frac{\log(1.0-u)}{a}\right]^{1/c},
\end{equation} 
where $u$ is here and hereafter the uniform random number at [0,1]; $M_{12}=M/(10^{12} M_\odot)$, $M_{min 12}$~--- lower cut-off of the mass spectrum. We get $a=0.07857$, $c=0.5822$, $M_{min 12}=50$. The minimal mass is taken  according to \citet{comerford2007}. The function (\ref{mass_function}) gives average and maximal fitting errors of only $0.35\%$ and $6.7\%$ relatively to the results by \citet{klypin2011}. In this paper all masses, except scaled by $10^{12} M_\odot$, are expressed in $h^{-1}$ units. The concentration function has also been adopted from \citet{klypin2011}, which is a function of a deflector redshift.  
In addition, some gaussian noise is added to relations in the cited above paper. The $3\sigma$ is set to give $40\%$ noise, where $\sigma$  is equal to the maximum uncertainty for the concentration parameter data in the cited above paper.  Density slope $\beta$ for Einasto model is varied as a function of mass defined by \citet{dutton2014}. 

Although, the 90\% of the cluster is assumed to be smooth mass distribution, the rest fraction is contributed by galaxies. Their masses are generated by the same law (\ref{mass_function}) with the following parameters: $M_{min 12}=0.001$, $a=1.57405$, $c=0.377378$ providing the mean and maximal relative errors of $2.6\%$ and $24\%$. Density model for each galaxy is spherically symmetric obeying Einasto law, having the same slope~--- mass function as applied for the cluster. In addition, for each galaxy a random uniform at [$-0.03$, $0.03$] noise is added,  whose range shows uncertainty in the fit of cosmological simulations \citep{dutton2014}. Concentration for galaxies is also calculated  in the same way as for the cluster with gaussian noise ($3\sigma$ equals to 40\%).  Throughout the simulation run, first the cluster model is created, then a test galaxy is placed randomly within it and the deflection quantities (deflection, its derivatives, shear and convergence) are calculated. The  position angle of the test galaxy is uniform, whilst the projected radius-vector is either specified explicitly or varied uniformly in [0, $0.25r_{vir}$] and [0, $0.5r_{vir}$]. The location of the other cluster galaxies are generated randomly according to the same density model as for the smooth component of the cluster.  

\section{Results}
The main results of the simulations are presented in tables and figures bellow. In order to guarantee stable statistical results when calculating the shear and convergence the model is run $10^6$ times once a parameter under study is changed. The influence of the following quantities and properties are investigated: cluster ellipticity, the impact of close galaxies and the deflector redshift. 

\begin{figure*}
\begin{center}
\includegraphics[scale=0.8]{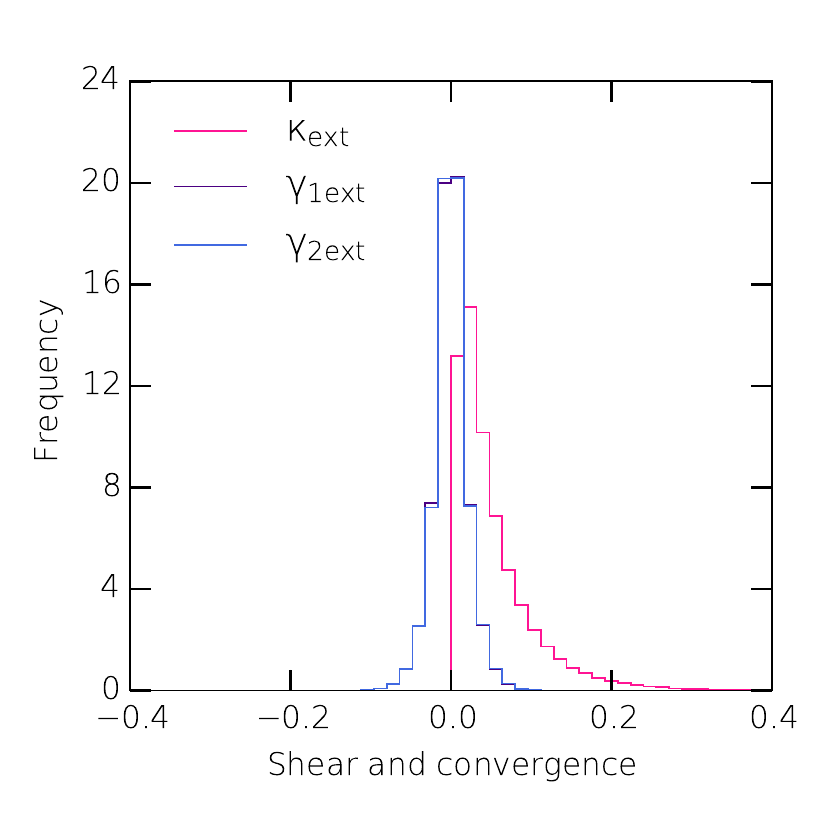}
\includegraphics[scale=0.8]{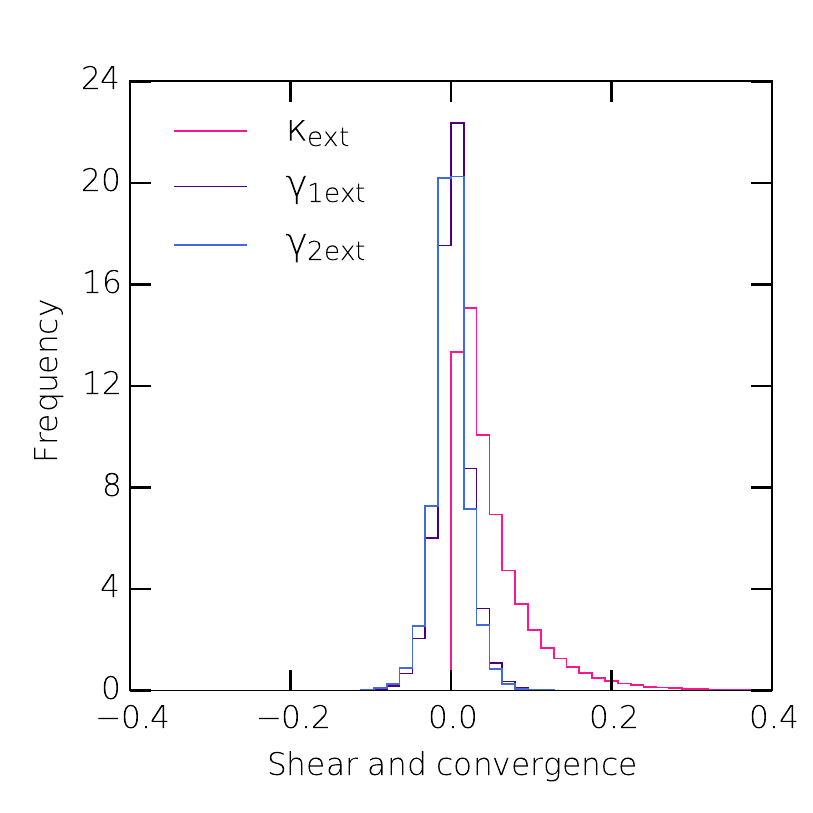}
\includegraphics[scale=0.8]{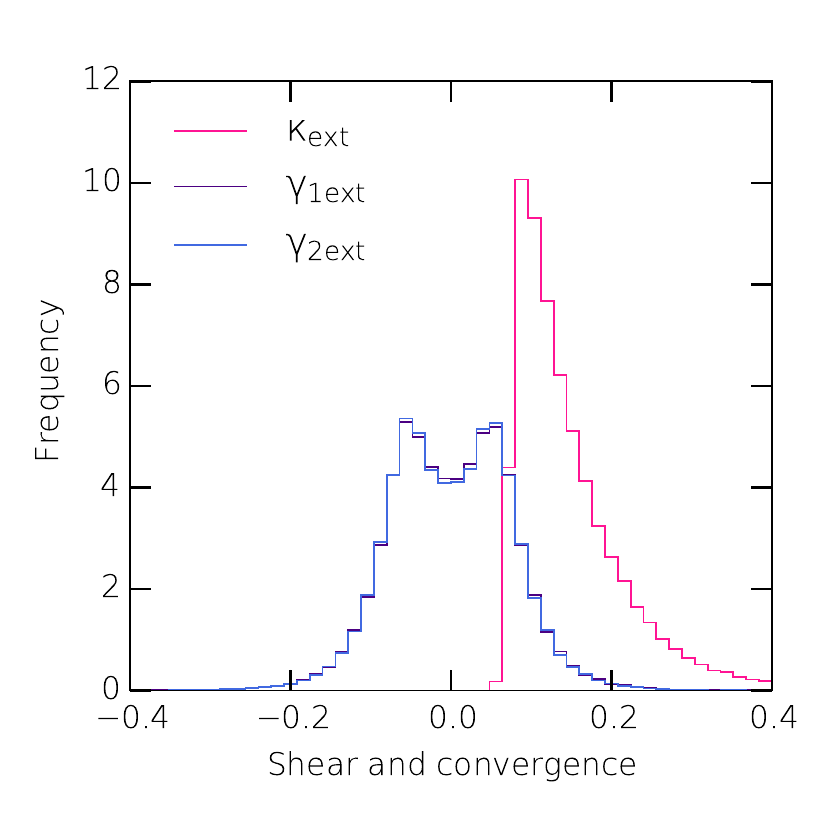}
\includegraphics[scale=0.8]{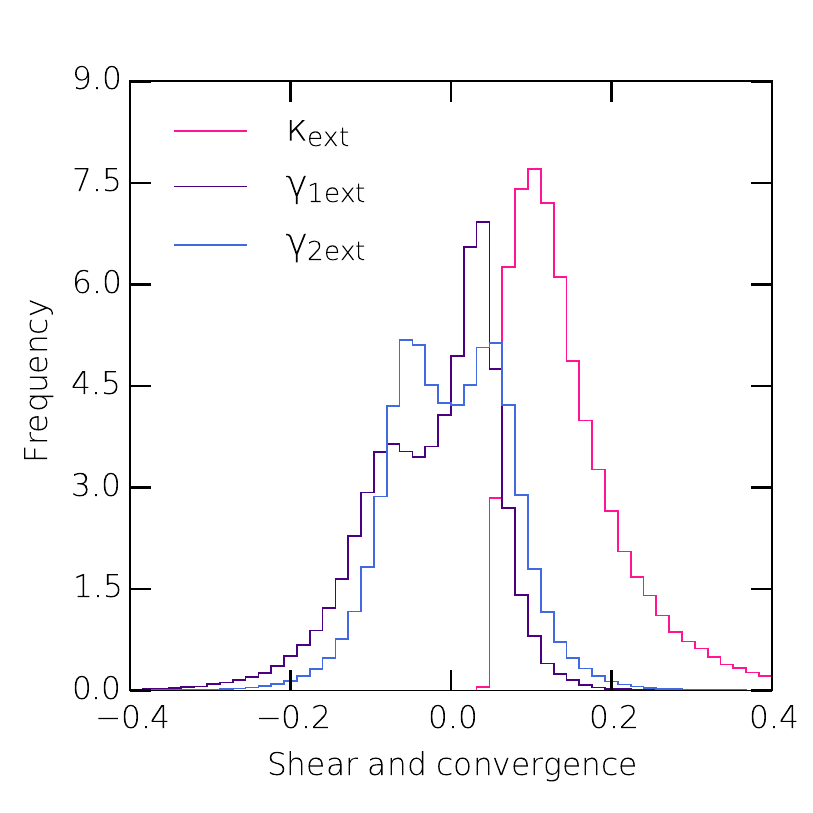}
\includegraphics[scale=0.8]{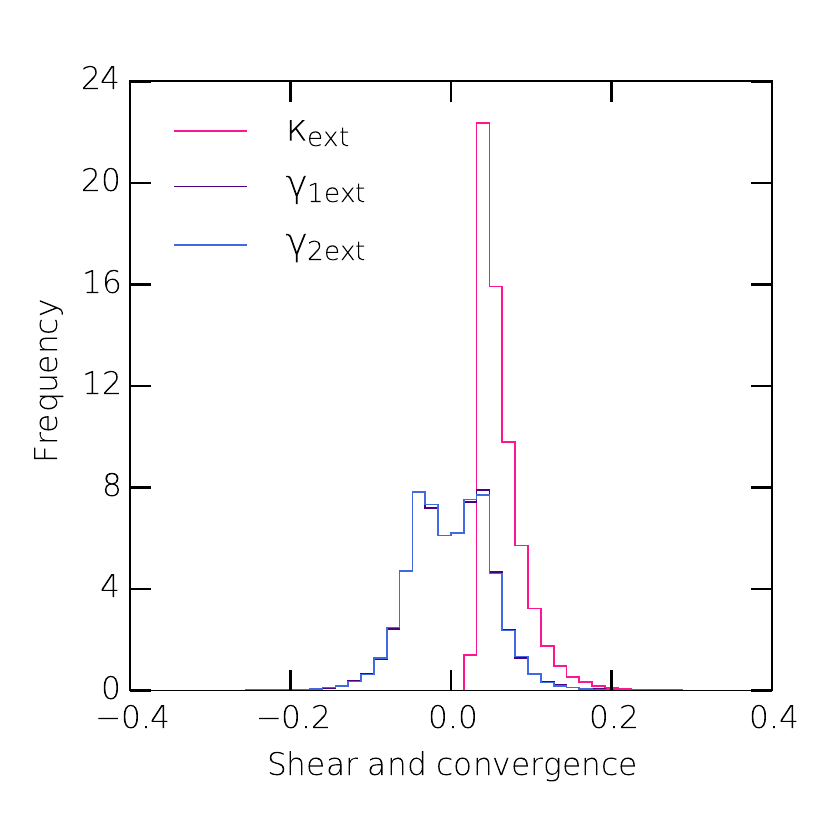}
\includegraphics[scale=0.8]{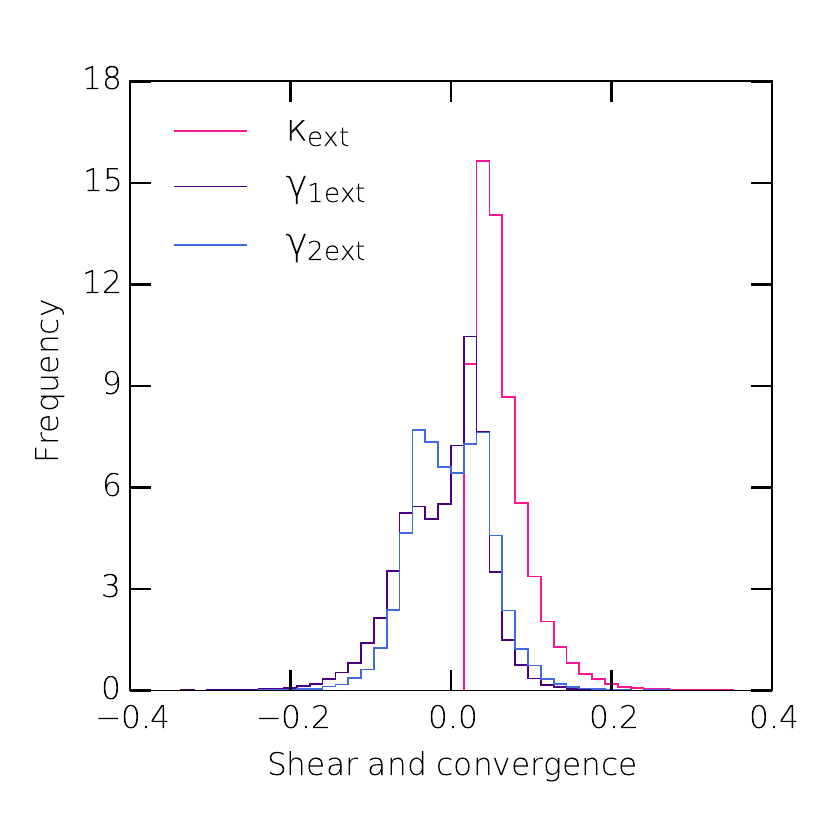}
\caption{External shear and convergence distribution for a galaxy at fixed distance from the centre of the galaxy cluster. \textit{Left column:} circular projected mass; \textit{right column:} elliptical mass with $q=0.75$. \textit{Top:} $r=0$; \textit{middle:} $r=0.25r_{vir}$; \textit{bottom:} $r=0.50r_{vir}$.}\label{basic}
\end{center}
\end{figure*}

\begin{figure*}
\begin{center}
\includegraphics[scale=0.8]{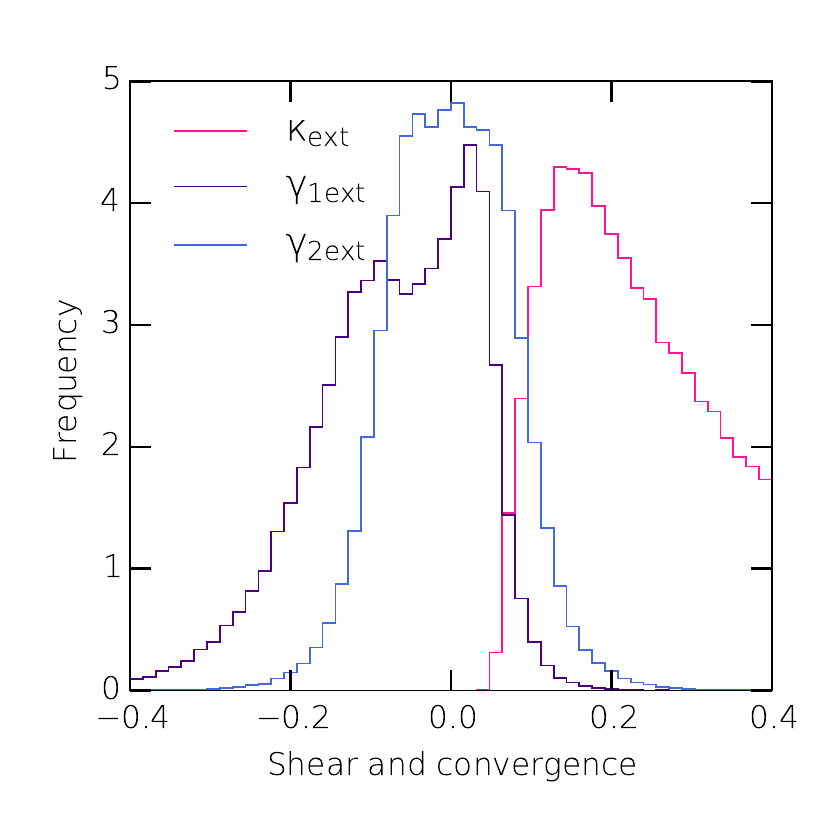}
\includegraphics[scale=0.8]{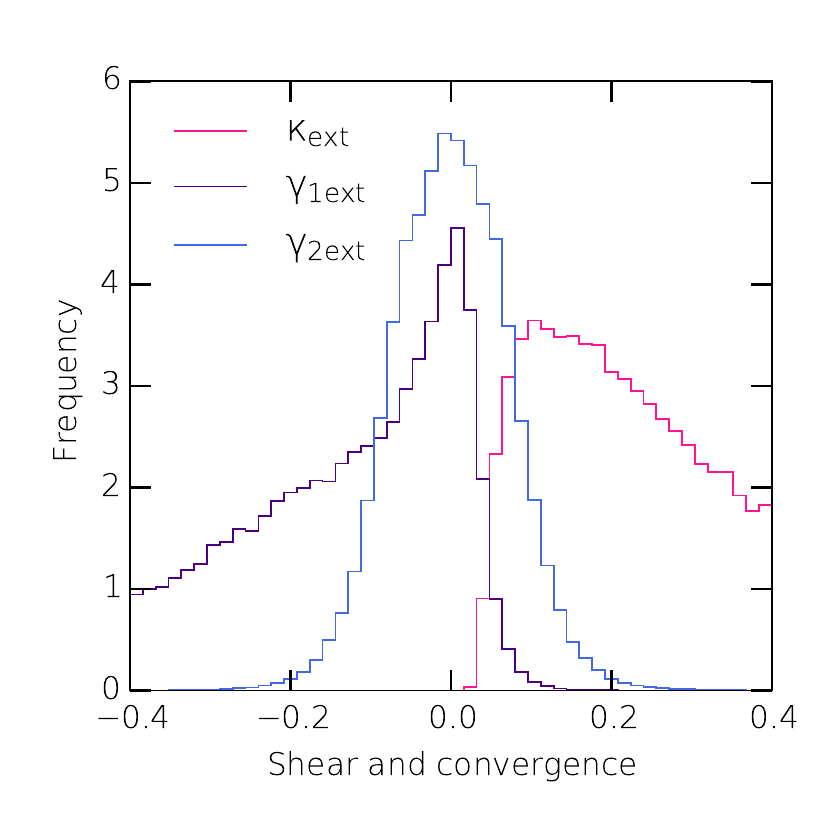}
\includegraphics[scale=0.8]{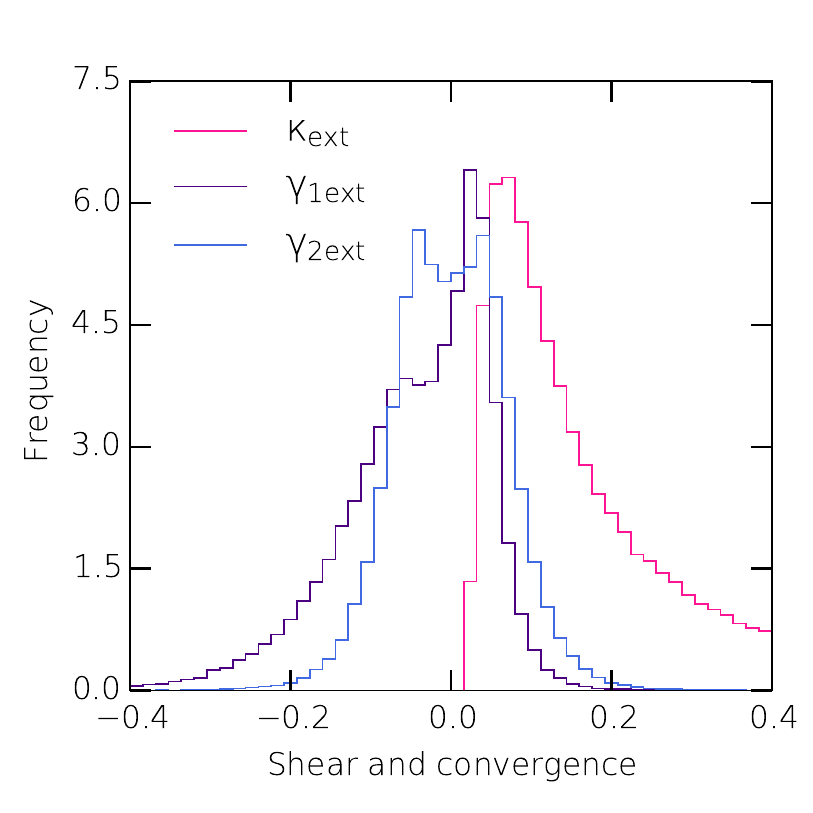}
\includegraphics[scale=0.8]{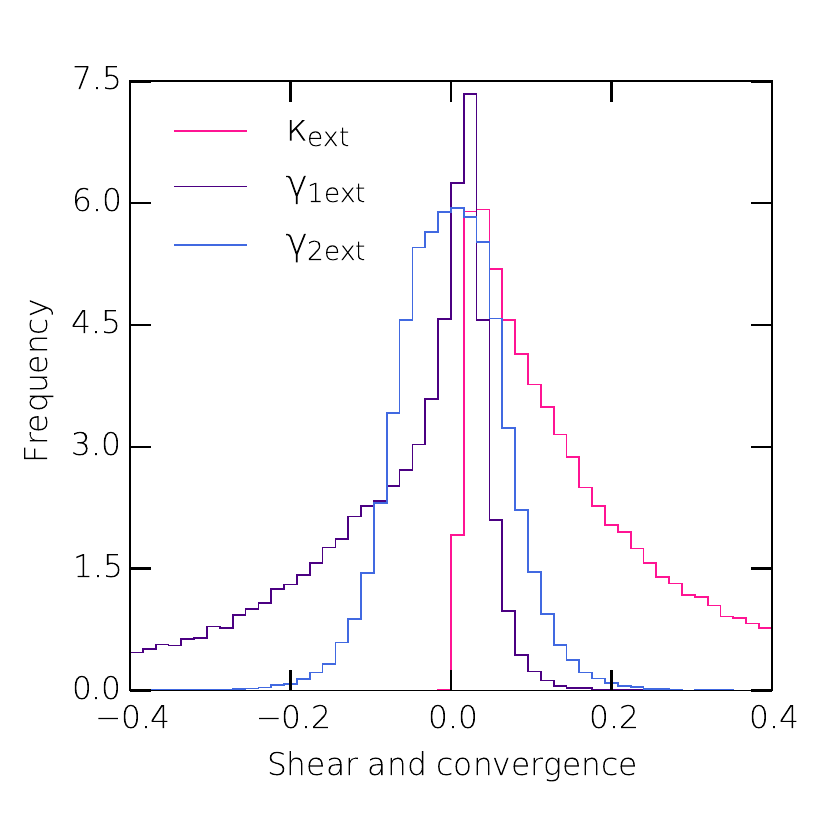}
\caption{Distribution of convergence and shear for a galaxy at various distances from the centre of the cluster. \textit{Left column:} $q=0.75$;  \textit{right column:} $q=0.50$. \textit{Top:} $r_{max}=0.25r_{vir}$; \textit{bottom:} $r_{max}=0.50r_{vir}$. } \label{centre}
\end{center}
\end{figure*}

The distributions of $\kappa_{ext}$, $\gamma_{1ext}$ and $\gamma_{2ext}$ for circular and some cases of elliptical mass projected mass are shown in figures~\ref{basic} and \ref{centre}. In histograms the horizontal axis represents dimensionless quantities of shear and convergence, the vertical axis shows normalised frequency of occurrences of a quantity.

For $10^6$ runs  histograms of $\gamma_{1ext}$ and $\gamma_{2ext}$ are practically indistinct. This can be explained by considering $\gamma_{1ext}$ and $\gamma_{2ext}$ for a circular case. From (\ref{derivatives}) it follows that ($\phi=\tan^{-1} (x_2/x_1)$):
\begin{eqnarray}
\gamma_{1ext} = \frac{1}{2}\cos(2\phi) \left[\frac{\partial \alpha}{\partial r} - \frac{\alpha}{r}\right], \\ \nonumber 
\gamma_{2ext} = \frac{1}{2}\sin(2\phi) \left[\frac{\partial \alpha}{\partial r} - \frac{\alpha}{r}\right].
\end{eqnarray}  

It is easily seen that for multiple simulations both gammas give the same average results. 

Table~\ref{table1} contains basic statistical data (mean, median and root mean square) for various cluster models. For comparative analysis the table contains cluster models having only smooth mass distribution as well as the including galaxies. These cases are marked in the table with ''on'' or ''off''. Mean and median for the data sets are defined in a standard way, the r.m.s. is calculated as a biased estimator.  All results are shown up to three decimal places as long as $N_{runs}$ approaches $\sim 10^6$ such a precision is sufficient. The largest value of the standard deviation in the simulations is $\sigma_{max}=0.277$, which corresponds to $\kappa_{ext}$ ($q=0.5$, smooth cluster mass). For $N=10^6$ this leads to the largest standard error of $s_{max} = \sigma_{max}/\sqrt{N}\simeq 2.8\times 10^{-4}$. For $3s_{max}$ we have $0.8\times 10^{-3}$ that corresponds to the three decimal places of the results. 

\begin{deluxetable}{ccccccccc}
\tabletypesize{\footnotesize}
\tablewidth{0pt}
\rotate
\tablecaption{Statistical results of Monte-Carlo simulations of the external shear and convergence for the test galaxy placed randomly in [0, $0.5r_{vir}$] in its galaxy cluster
\label{table1}}
\startdata
\hline\hline
 & & & & & & & & \\
Number of runs & Axes ratio & Nearby galaxies & Mean($\kappa_{ext}$) & Median($\kappa_{ext}$) & 
R.M.S.($\kappa_{ext}$) & Mean($\gamma_{ext}$) & Median($\gamma_{ext}$) & R.M.S.($\gamma_{ext}$) \\
& & & & & & & & \\
\hline%
 $10^6$  & $1.00$     & OFF & 0.211  & 0.130 & 0.214 &  0.093 & 0.084 & 0.047 \\
 $10^6$  & $1.00$     & ON   & 0.218  & 0.140 & 0.210 &  0.088 & 0.078 & 0.047 \\ 
 $10^6$  & $0.90$     & OFF & 0.212   & 0.131 & 0.214 & 0.096  &  0.085 & 0.050 \\ 
 $10^6$  & $0.90$     & ON   & 0.218   & 0.142 & 0.211 & 0.090 &   0.080  & 0.048 \\ 
 $10^6$  & $0.80$     & OFF & 0.213   & 0.131 & 0.220 & 0.104 &   0.087  & 0.064 \\ 
 $10^6$  & $0.80$     & ON   & 0.219   & 0.141 & 0.216 & 0.097 &   0.082  & 0.060 \\ 
 $10^6$  & $0.70$     & OFF & 0.218   & 0.133 & 0.232 & 0.118 &   0.090  & 0.089 \\ 
 $10^6$  & $0.70$     & ON   & 0.222   & 0.143 & 0.222   & 0.109   &   0.085  & 0.080 \\ 
 $10^6$  & $0.60$     & OFF & 0.224   & 0.136 & 0.245   & 0.135   &   0.093  & 0.122 \\ 
 $10^6$  & $0.60$     & ON   & 0.228   & 0.143 & 0.237   & 0.124   &   0.088  & 0.109 \\ 
 $10^6$  & $0.50$     & OFF & 0.236   & 0.137 & 0.277   & 0.162   &   0.097  & 0.171 \\ 
 $10^6$  & $0.50$     & ON   & 0.236   & 0.144 & 0.259   & 0.145   &   0.090  & 0.148 \\ 
\enddata
\end{deluxetable}

\begin{figure*}
\begin{center}
\includegraphics[scale=1]{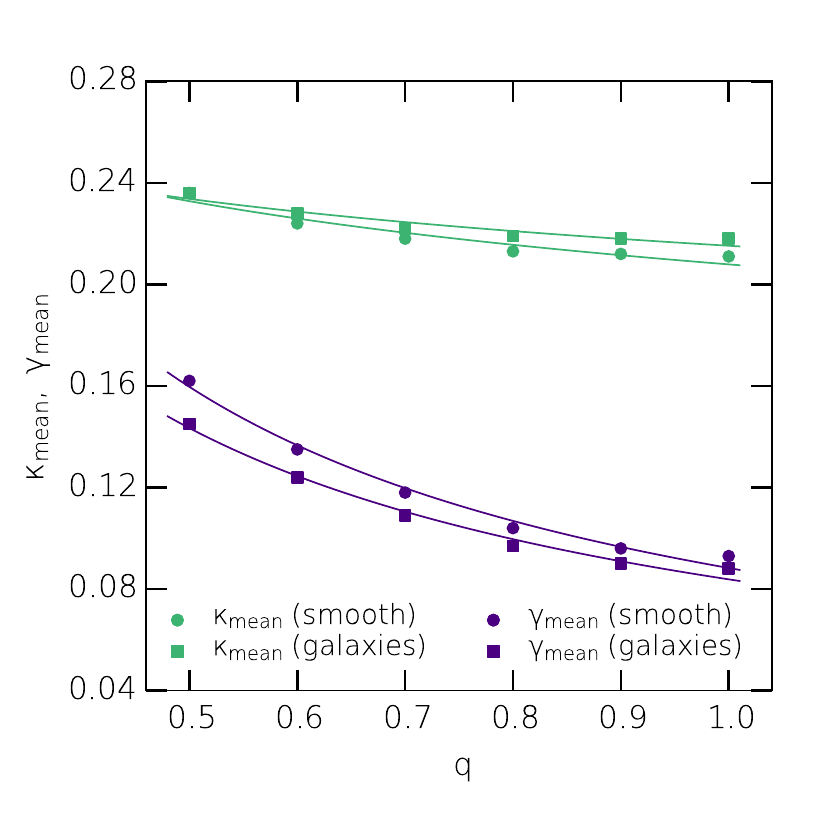}
\caption{Distribution characteristics of convergence and shear as a function of cluster ellipticity. Curves represent $A/q^B$ power fits.} \label{ellipticity}
\end{center}
\end{figure*}

\begin{figure*}
\begin{center}
\includegraphics[scale=1]{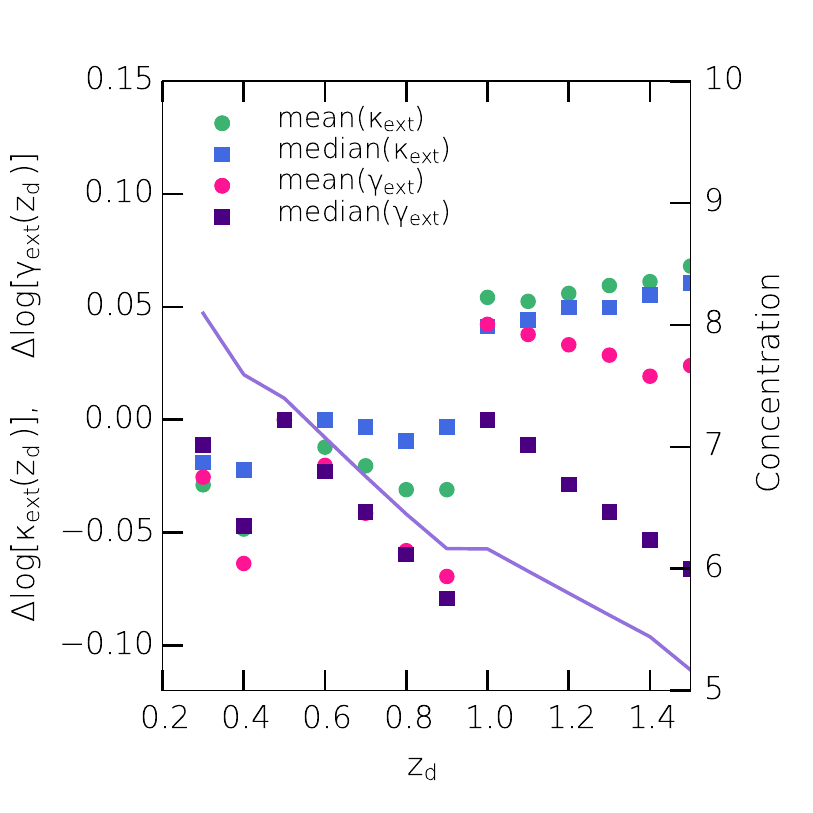}
\caption{Distribution characteristics of convergence and shear as a function of deflector redshifts. The ratio $D_dD_{ds}/D_s$ is fixed throughout the simulations. Spherically symmetric cluster model is applied. $\Delta\log x_d\equiv \log x_d - \log x_0$, where $x=\kappa_{ext}, \ \gamma_{ext}$. Initial values correspond to $z_d=0.5$ and $z_s=1.5$.}\label{redshift}
\end{center}
\end{figure*}

The influence of the cluster ellipticity and the deflector redshift on $\kappa_{ext}$ and $\gamma_{ext}$ are shown in figures \ref{ellipticity} and \ref{redshift} respectively. For ellipticity study axial ratio is set $q=0.75\ldots 1.0$. Curves in figure \ref{ellipticity} show gradual increase. 

It is worth interest to study redshift dependence as the concentration parameter both for the cluster and galaxies depends on the deflector redshift. To avoid angular size distance dependence, the ratio $D_d D_{ds}/D_s$ is kept fixed, the same as for $z_d=1$ and $z_s=3$. New values of the deflector and source redshifts are chosen in order to maintain this ratio. 
Results of redshift dependence of the shear and convergence could be potentially compared with anticipated data from ongoing and future observations.        

\subsection{Lensing system IRC0218}
\begin{figure*}
\begin{center}
\includegraphics[scale=1]{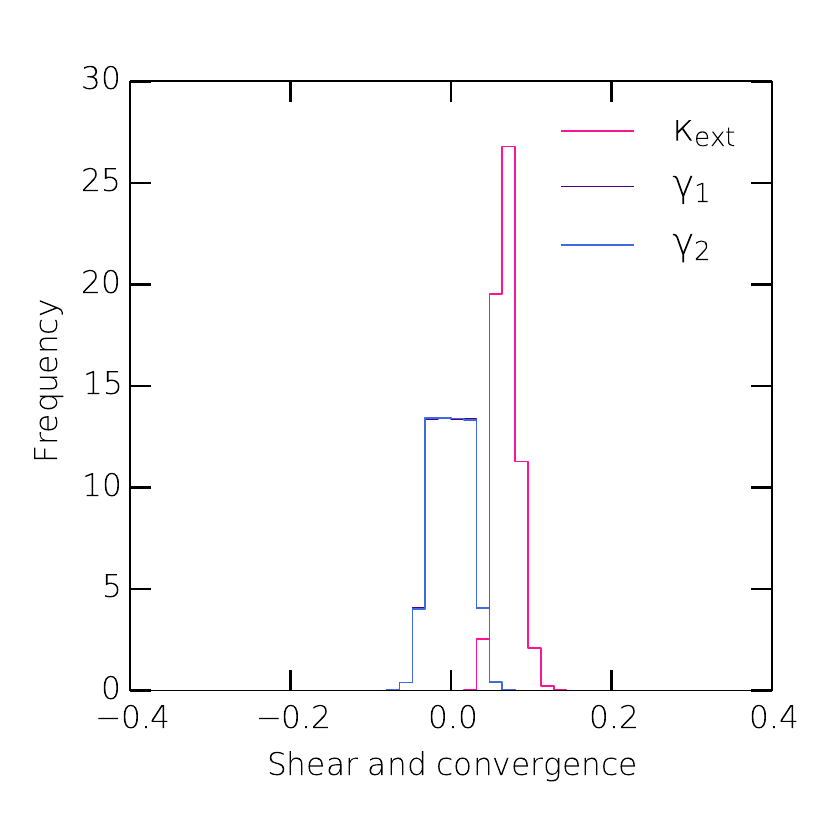}
\includegraphics[scale=1]{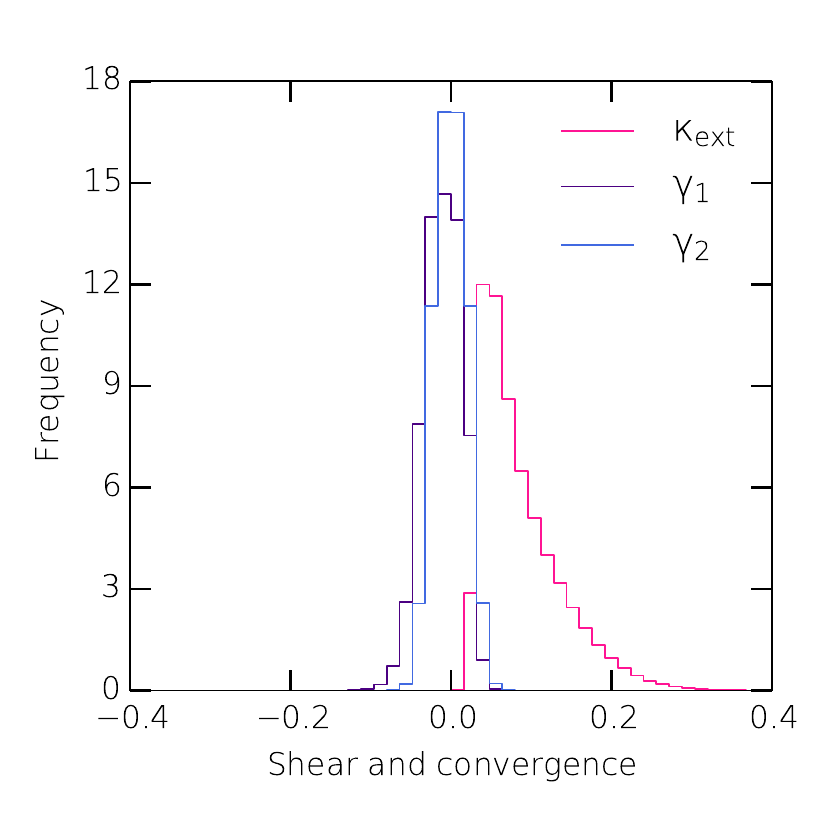}
\caption{Distribution of convergence and shear for a strong lens in the cluster IRC-0218. \textit{Top:} Circular projected cluster model, fixed position of a lens galaxy at $13.3''$. \textit{Bottom:} Variable position of the galaxy from 0 to 25 arcsec from the cluster's centre, uniform ellipticity from 0.7 to 1.0.} \label{irc0218}
\end{center}
\end{figure*}

I apply the cluster model simulate the convergence, shear and magnification caused by a cluster environment influenced on a strong lens seen through this cluster \citep{pierre2012,wong2014}. The following data are used for the Monte Carlo simulations: $z_d=1.62$, $z_s=2.26$; cluster mass $M=(7.7\pm 3.8)\times 10^{13}M_{\odot}$, position of lens relatively to the cluster centre $r_d=13.3''$, uncertainty of the cluster centre location $\Delta r_c=25''$. Also, here I adopt the same cosmological parameters as used by \citet{wong2014} in order get comparable results: $H_0=70$~km/s/Mpc, $\Omega_\Lambda=0.70$, $\Omega_m=0.30$. The results are presented in the table~\ref{table2} and figure~\ref{irc0218}. 

\begin{deluxetable}{ccccccccc}[h]
\tabletypesize{\footnotesize}
\tablewidth{0pt}
\tablecaption{External shear and convergence for the cluster IRC-0218
\label{table2}}
\startdata
\hline\hline
\# & Number of runs & Mean($\kappa_{ext}$) & Median($\kappa_{ext}$) & 
R.M.S.($\kappa_{ext}$) & Mean($\gamma_{ext}$) & Median($\gamma_{ext}$) & R.M.S.($\gamma_{ext}$) \\
\hline%
1 & $10^6$  & 0.0862  & 0.0713 & 0.0494   &  0.0252 & 0.0252 & 0.0114 \\
2 & $10^6$  & 0.0698  & 0.0689 & 0.0137 &  0.0301 & 0.0292 & 0.0102 \\ 
3 &  $10^6$ & 0.0869   & 0.0721 & 0.0506 & 0.0289  &  0.0272 & 0.0141 \\ 
\enddata

1~--- spherical cluster, variable lens position from 0 to 25 arcsec from the cluster centre; 2~--- spherical cluster, exact position of the lens $13.3''$ from the cluster centre; 3~--- variable lens position from 0 to 25 arcsec, variable cluster ellipticity from 1.0 to 0.7.
\end{deluxetable}

Therefore, using average values of $\kappa_{ext}$ and $\gamma_{ext}$ and substituting them to the standard definition of magnification, we have  $\mu=1.2$. 

\section{Discussion and Conclusion}

Based on the results of Monte Carlo simulations the main conclusions of the research are the following. 

Throughout the simulations parameters are varied to obtain  comprehensive estimations of the external shear and convergence. Einasto spatial density profile has proven itself as an applicable for modelling clusters.

For moderate values of axial ratio ($q=0.5\ldots 1.0$) the mean external shear is bounded at $0.21\ldots 0.24$, whereas the mean shear is $0.1\ldots 0.2$. Both quantities could rise significantly if the cluster has considerable ellipticity or a test galaxy is located near the cluster centre. Figure~\ref{redshift} shows significant deviations of $\kappa_{ext}$ and $\gamma_{ext}$ when the deflector redshift changes, that it related to the concentration dependence on the redshift.    

The results reveal a well noticeable impact of close galaxies treated separately (table~\ref{table1}) rather than including them into smooth mass component. In this case, the mean and median values are systematically larger without taking into account individual galaxies. The differences considerably exceed the precision threshold of the simulated data.  

Application of the model to the cluster IRC-0218 allows to get reliable statistical estimations for $\kappa_{ext}$, $\gamma_{ext}$ and magnification. The latter parameter is in a good correspondence with estimations of external mass by \citet{wong2014}.   

In this article the first thorough calculations of external convergence and shear caused by a galaxy cluster environment has been worked out. In most other relevant publications \citep{wambsganss1999, metcalf2012, xu2013} such calculations are either omitted or estimated roughly. However, the proper extraction of various deflecting effects is essential for searching fine imprints of lensing by subhaloes of galaxies.  

\bibliographystyle{apj}
\bibliography{lensing}

\end{document}